\documentstyle[nato,numreferences,epsfig]{crckapb} 

\newcommand{\mb}[1]{ {\mbox{\boldmath{$#1$}}}  }

\begin{opening}
\title{S- AND D-WAVE PAIRING IN SHORT COHERENCE LENGTH SUPERCONDUCTORS}
\subtitle{}

\author{James F. Annett and J.P. Wallington}
\institute{University of Bristol,\\
 H.H. Wills Physics Laboratory, \\ Royal Fort, Tyndall Avenue,\\
 Bristol BS8 1TL, UK.}
\end{opening}

\runningtitle{S-WAVE, D-WAVE AND MIXED PAIRING STATES}

\begin{document}

% The \begin{document} command comes after the \end{opening}

\section{Introduction}

Over the past three years or so the evidence has become overwhelming
that the cuprate high T$_c$ superconductors have a $d$-wave pairing
state\cite{ginsberg}.  The measurements of Wollman {\it
et al.}\cite{wollman} and of Tsuei {\it et al.}\cite{tsuei} are especially
convincing since they do not depend on the microscopic physics of the
energy gap, but instead depend only on the order parameter phase.
Other experiments, such as photoemission\cite{ding,shen} and the
temperature dependence of the penetration
depth\cite{bonn,cooper,waldram}, also strongly support the $d$-wave
picture.
\\ \\
On the other hand, there is continuing controversy over whether the
pairing state is a pure $d$-wave or an $s-d$
mixture\cite{onellion,klemm}.  There is indeed evidence for a
significant $s$-wave component in YBa$_2$Cu$_3$O$_7$\cite{dynes}.  A
subdominant $s$-wave component could be compatible with the Wollman {\it
et al.} and the Tsuei {\it et al.}  experiments provided that it was
not too large.  The photoemission and penetration depth measurements
also cannot rule out a small $s$-wave component (either $s \pm d$ or
$s \pm id$) , although they can possibly put upper bounds on the magnitude
of the $s$ component.
\\ \\
Any observations of an $s$ component have important implications for
the various theories of the pairing mechanism.  For example,
antiferromagnetic spin fluctuations lead to attraction in the
$d_{x^2-y^2}$ pairing channel, but are pair breaking in the $s$-wave
channel\cite{afm}. Similarly the Hubbard model with a positive on-site
interaction U may have a $d_{x^2-y^2}$ paired ground state, but would
presumably not support $s$-wave Cooper pairs.  In the case of
YBa$_2$Cu$_3$O$_7$ the orthorhombic crystal symmetry makes some
non-zero $s$-wave component inevitable, but a large $s$-wave component
would be difficult to reconcile with either of these pairing
mechanisms. On the other hand, pairing mechanisms based on
electron-phonon interactions, polarons, or other non-magnetic
excitations (e.g. excitons, acoustic plasmons) could be compatible
with either $s$-wave or $d$-wave pairing states\cite{varegiolanis}.
In these models, whether $s$-wave, $d$-wave or mixed pairs are more
strongly favoured would depend on details of the model parameters, and
could even vary from compound to compound.  Indeed there is some
evidence that the n-type cuprate superconductors are
$s$-wave\cite{anlage} (or at least they have no zeros in the gap
$|\Delta({\bf k})|$ on the Fermi surface).  This would imply that
either the pairing mechanism is different for the n- and p-type
materials, or that the mechanism allows both $s$-wave or $d$-wave
ground states depending on the band filling.
\\ \\
In this paper we examine the attractive nearest neighbour Hubbard
model, which is the simplest model that allows $s$-wave, $d$-wave or
mixed pairing states to occur\cite{ranninger}.  We examine the overall
phase diagram, paying particular attention to the regions near the
phase transitions between the $s$, $d$ and $s \pm id$ phases.  We derive
an appropriate Landau-Ginzburg-Wilson (LGW) effective action for the
model. In the mean-field approximation this LGW functional reproduces
the usual Hartree-Fock-Gor'kov equations.  Beyond mean field theory
this functional allows us to examine the large interaction limit, in
which the superconducting phase transition becomes a Bose-Einstein
condensation of preformed pairs.  In this limit there will be a
pseudogap in the normal state density of states.  The strong coupling
limit is especially interesting in this model because, unlike the case
of on-site interactions, there are at least two `species' of preformed
pairs, $d_{x^2-y^2}$ and $s$.  We show below that these become
degenerate in the strong coupling Bose-Einstein limit.

\section{The Nearest Neighbour Attractive Hubbard Model}

Our starting point is the nearest-neighbour attractive
Hubbard model in two dimensions,
\begin{equation}
\hat{H} =  - t \sum_{ \langle ij\rangle \sigma}
 \left( c^\dagger_{i\sigma} c_{j\sigma} +H.C. \right)
 - V \sum_{\langle ij \rangle } 
 n_{i}n_{j},  \label{ham}
\end{equation} 
where $\sum_{\langle ij \rangle}$ denotes summation over all of the
bonds between the nearest neighbour sites. For simplicity
we ignore any on-site interaction terms. 
\\ \\
The mean-field gap equations can be used straightforwardly to
calculate T$_c$ for this model (Section 4). The gap equation has
stable solutions for (extended) $s$-wave, $p$-wave, and $d$-wave
pairing states, depending on the model parameters. One can see in
Fig. 1(a)-(c) that, for all $V$, 
\pagebreak
\begin{figure}[h!]
\centerline{\epsfig{file=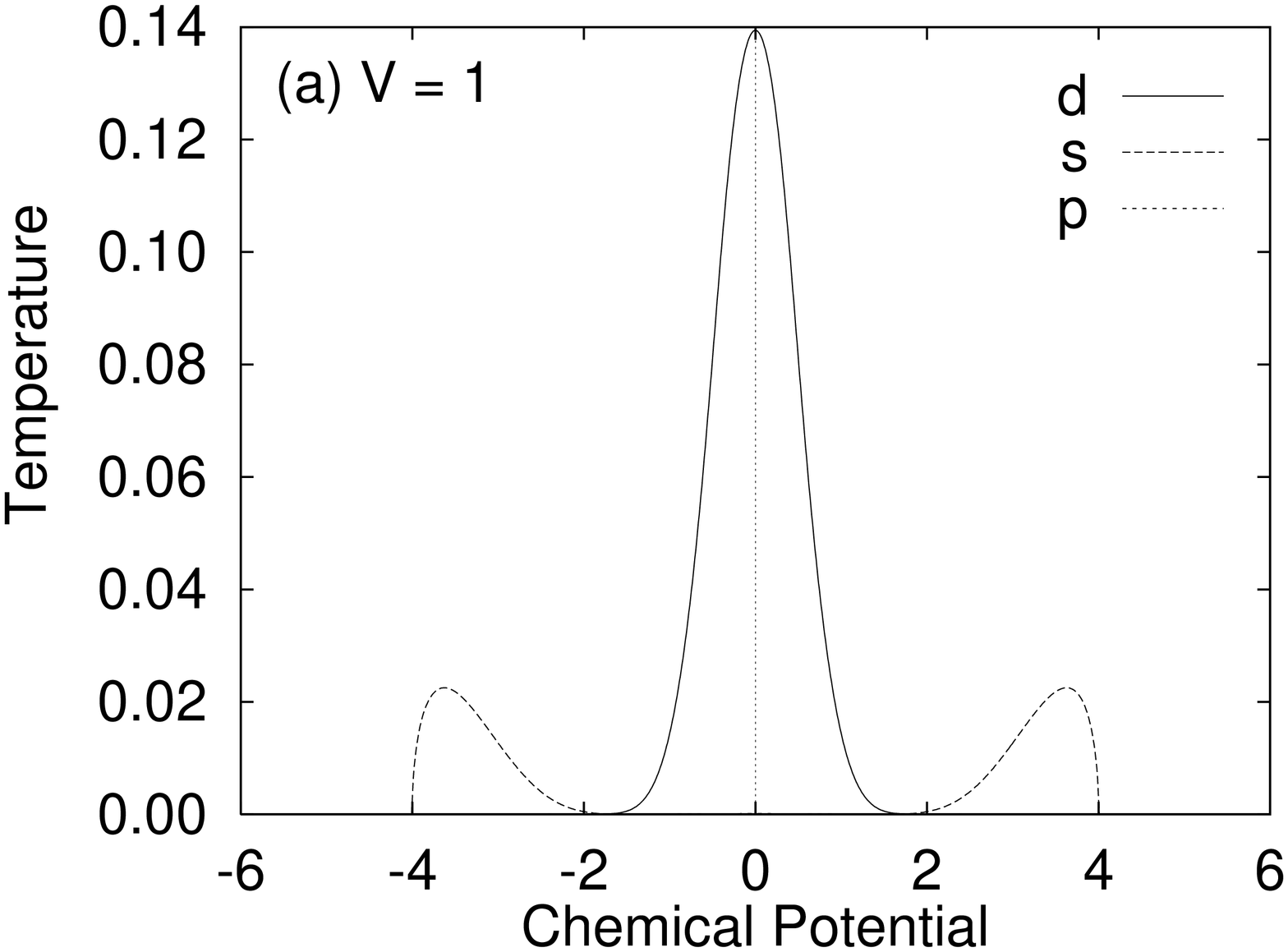,width=5.8cm,angle=-90}}
\centerline{\epsfig{file=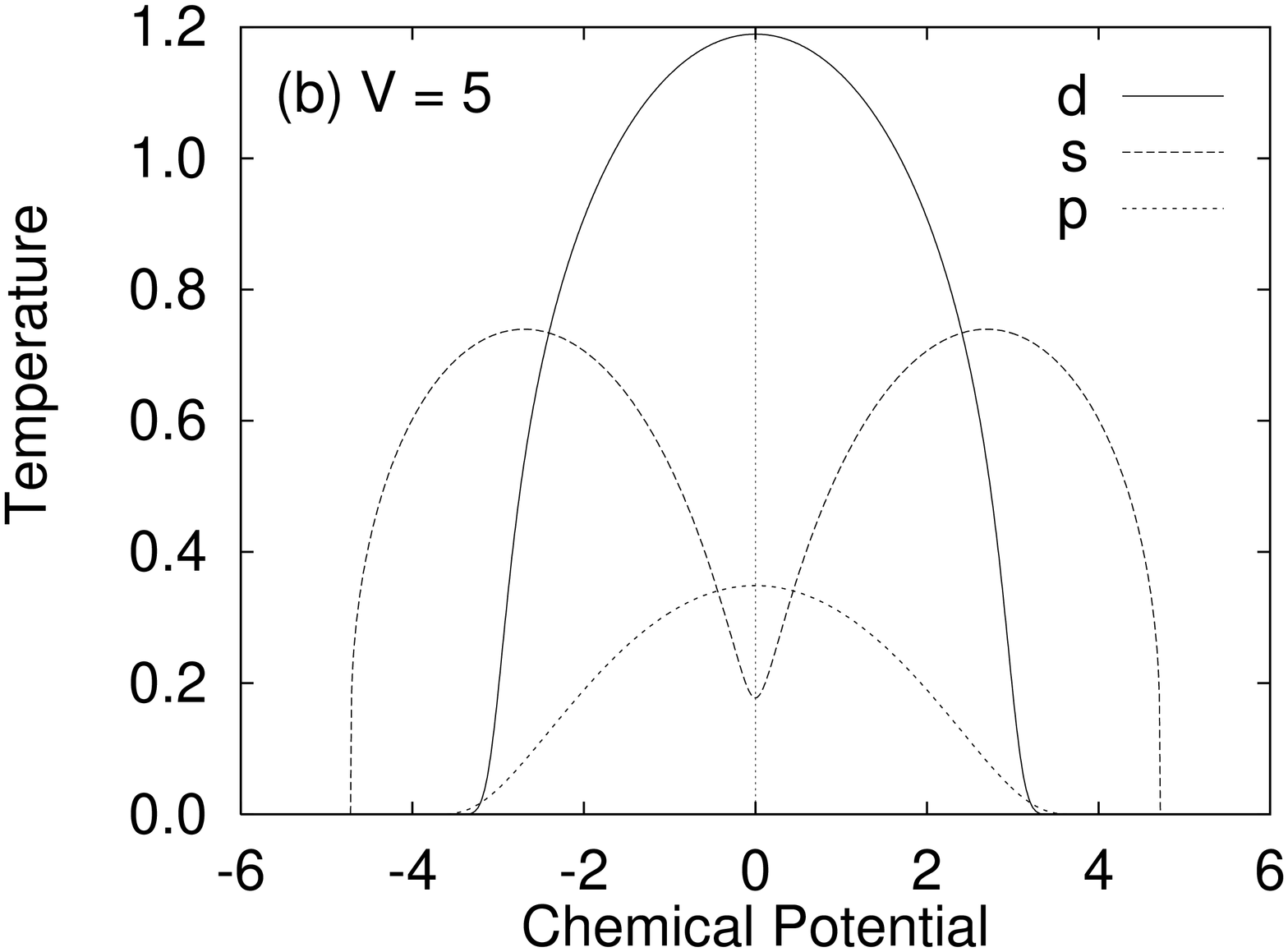,width=5.8cm,angle=-90}}
\centerline{\epsfig{file=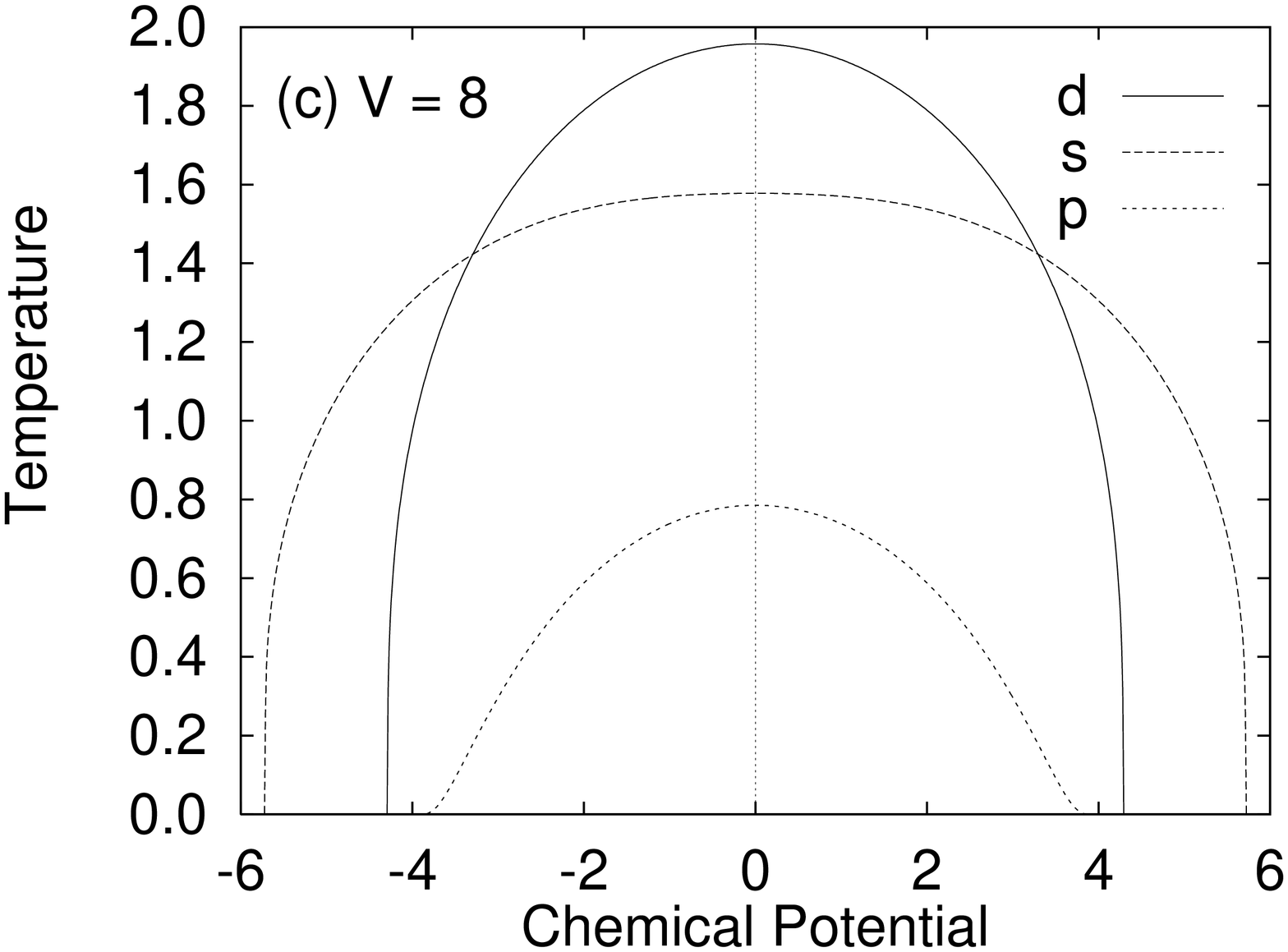,width=5.8cm,angle=-90}}
\caption{Mean-field transition temperature, T$_c$,  versus chemical potential, $\mu$, for $s$-, $p$- and $d$-wave superconductivity at different interaction strengths; (a) $V=1t$, (b) $V=5t$ and (c) $V=8t$.}
\label{figure1}
\end{figure}
\pagebreak
the $d$-wave state is strongly favoured at and near to the van Hove
singularity at $\mu = 0$, while the $s$-wave state is favoured in the
limits of nearly full or empty bands.  The $p$-wave solution also has
a maximum at $\mu=0$ but is always small compared with the $d$-wave.
The mean-field phase diagram also includes a small region of $s \pm id$
pairing (not shown in Fig. 1) just below the points where the $d$- and
$s$-wave T$_c$ cross.  These crossing points are therefore
tetracritical points, where all four phases (normal, $s$, $d$ and
$s \pm id$) meet.  Below we shall closely examine the nature of the phase
diagram in the vicinity of these tetracritical points.
\\ \\
If next nearest neighbour hopping terms are also included in the
Hamiltonian then there is no longer particle-hole symmetry about
half-filling and and it is possible to have $d$-wave pairing
predominantly for p-type doping and $s$-wave pairing predominantly for
n-type doping. This would be qualitatively consistent with the
experimental evidence for $s$-wave pairing in the n-type
cuprates\cite{anlage}.  However, for simplicity, below we shall
concentrate on the case of nearest-neighbour hopping only.

\section{The Landau-Ginzburg-Wilson Effective Action}

In order to examine how fluctuations modify the mean-field picture of
Fig. 1 is is useful to develop a Landau-Ginzburg-Wilson (LGW)
effective action.  This allows us to examine both the weak coupling,
$V\ll8t$, and the strong coupling, $V\gg8t$, limits on an equal footing.
Only in the strong or intermediate coupling regime will fluctuation
effects lead to phenomena such as the normal-state pseudogap which is
observed in several different high T$_c$ materials\cite{pseudogap}.
\\ \\
The Landau-Ginzburg-Wilson effective action of the nearest-neighbour
Hubbard model can be derived in a way similar to the usual $s$-wave
case of on-site interaction\cite{popov}.  Firstly the grand canonical
partition function is written:
\begin{eqnarray}
{\cal Z} &=& 
   \int {\cal D}\big[c,c^{\dagger}] 
        e^{{\cal S}},\\
{\cal S} &=& 
   \int_{0}^{\beta} {\rm d}\tau 
   \left[
   \sum_{i\sigma} c^{\dagger}_{i\sigma}(\tau) \partial_{\tau}
 c_{i\sigma}(\tau)  
   -\hat{H}_0 -\hat{H}_I
   \right],
\end{eqnarray} 
where $\hat{H}_0$ is given by the hopping term in the Hamiltonian
Eq. \ref{ham} (including a chemical potential term, $-\mu\hat{N}$) and
$\hat{H}_I$ is the interaction term.
\\ \\
The interacting part may be neatly written in terms of pairing
operators on nearest neighbour sites:
\begin{eqnarray}
\hat{H}_I &=& -V\sum_{\langle ij \rangle}
               {\rm tr}\left(F^{\dagger}_{ij}F_{ij}\right),
\end{eqnarray}
where $F_{ij}$ is defined by:
\begin{equation}
F_{ij} = \left( \begin{array}{lr}
 c_{j\uparrow} c_{i\uparrow} & c_{j\downarrow} c_{i\uparrow}\\
 c_{j\downarrow} c_{i\uparrow} & c_{j\downarrow} c_{i\downarrow}
\end{array} \right) .
\end{equation}
The physical nature of the interaction is best illustrated by introducing
a new set of operators,
\begin{eqnarray}
B_{ij}^{00}       &\equiv&   \frac{1}{\sqrt{2}}
                        \left(
                        c_{j\uparrow}c_{i\downarrow}
                        -c_{j\downarrow}c_{i\uparrow}
                        \right),\\
B_{ij}^{11}       &\equiv&   c_{j\uparrow}c_{i\uparrow} ,\\
B_{ij}^{10}       &\equiv&   \frac{1}{\sqrt{2}}
                        \left(
                        c_{j\uparrow}c_{i\downarrow}
                        +c_{j\downarrow}c_{i\uparrow}
                        \right),\\
B_{ij}^{1\bar{1}} &\equiv&   c_{j\downarrow}c_{i\downarrow}.
\end{eqnarray}
In terms of these, $F_{ij}$ may be decomposed as
\begin{eqnarray}
F_{ij} &=& \frac{i\sigma_y}{\sqrt{2}}
           \left(  
           B_{ij} - \mb{\sigma}\cdot{\bf B}_{ij}
           \right),
\end{eqnarray}
where $\mb{\sigma}$ is the vector of Pauli matrices, and
\begin{eqnarray}
B_{ij}       &\equiv& B_{ij}^{00},\\
{\bf B}_{ij} &\equiv& \left(
         \frac{B_{ij}^{1\bar{1}}-B_{ij}^{11}}{\sqrt{2}},
         \frac{B_{ij}^{1\bar{1}}+B_{ij}^{11}}{-i\sqrt{2}},
         B_{ij}^{10}
                 \right).
\end{eqnarray}
The scalar, $B_{ij}$, is even under both time reversal and parity
whilst the vector, ${\bf B}_{ij}$, is odd under both.  We interpret
$B_{ij}$ and ${\bf B}_{ij}$ as the annihilation operators for singlet
and triplet Cooper pairs on the bond $\langle ij \rangle$\cite{volovik}.  
\\ \\
In terms of these operators the interaction Hamiltonian becomes:
\begin{eqnarray}
\hat{H}_I &=& -V\sum_{<ij>}
               {\rm tr}\left(F^{\dagger}_{ij}F_{ij}\right),  
               \nonumber \\
          &=& -V\sum_{<ij>}
               \left(
                    B_{ij}^{\dagger}B_{ij}
                   +{\bf B}_{ij}^{\dagger}\cdot{\bf B}_{ij}
               \right), \label{hamtwo} \label{hami}
\end{eqnarray}
where we have used tr$(\sigma_i)=0$ and
tr$(\sigma_i\sigma_j)=2\delta_{ij}$.  
\\ \\
Since this is a sum of squared bilinear Fermi operators, the
Hubbard-Stratonovi\v{c} transformation \cite{popov} can be employed.
The Gaussian identity,
\\ \\\begin{equation}
  e^{+VA^\dagger A} = \frac{V}{2\pi i} \int d\phi d\phi^*
 e^{-V\left(|\phi|^2 + A\phi^* + A^\dagger \phi\right)},
\end{equation}
where $A$ is a bilinear Fermi operator and $\phi$ is a c-number,
allows us to decouple the quartic interaction term in terms of
new bosonic fields. In the case of Eq. \ref{hami} this is accomplished
by introducing a
complex scalar field, $\psi_{ij}(\tau)$, and a complex vector field,
$\mb{\Psi}_{ij}(\tau)$, for each bond in the lattice and every
imaginary time.  The partition function can now be expressed
in terms of these Bose fields:
\begin{equation}
  {\cal Z} \equiv \int {\cal D}[\psi,\psi^*;\mb{\Psi},\mb{\Psi}^*]
  e^{S_{b}}. \label{eqz}
\end{equation}
The effective action is
\begin{equation}
  S_{b}  \equiv
 - V \int_{0}^{\beta} {\rm d}\tau \sum_{\langle ij \rangle }
      \left( |\psi_{ij}(\tau)|^2 + |\mb{\Psi}_{ij}(\tau)|^2 \right)
  + \ln{\int {\cal D}[c,c^\dagger]e^{{\cal S}_f}},
\end{equation}
with the remaining fermions contained in
\begin{eqnarray}
{\cal S}_f &\equiv&
    \int_{0}^{\beta} {\rm d}\tau 
    \Big[
    \sum_{i\sigma} 
        c^{\dagger}_{i\sigma}(\tau) \partial_{\tau}c_{i\sigma}(\tau)  
    -\hat{H}_0 \label{eqsf}
      \\
&\;&\qquad\qquad
    - V\sum_{\langle ij \rangle }
    \left(
    \psi_{ij}^*(\tau) B_{ij}(\tau) 
    +\mb{\Psi}_{ij}^*(\tau)\cdot{\bf B}_{ij}(\tau) + H.C. \right)
    \Big].\nonumber
\end{eqnarray}
Equations \ref{eqz}-\ref{eqsf} are a formally exact representation
of the nearest neighbour attractive Hubbard model.
\\ \\
The bosonic fields introduced above are defined separately for each
bond $\langle ij \rangle $ in the lattice. 
It is more convenient to
form site-centred combinations with a definite symmetry.
We define two singlet fields at site ${\bf r}_i$
\begin{eqnarray}
	\psi_s &    = &
\frac{1}{2} (\psi_x + \psi_{-x} + \psi_y + \psi_{-y}), \nonumber\\
        \psi_d  &  = & 
\frac{1}{2}(\psi_x + \psi_{-x} - \psi_y - \psi_{-y}),
\end{eqnarray}
where $\psi_a({\bf r}_i,\tau)\equiv\psi_{ii+a}(\tau)$ for $a=\pm x,\pm
y$ according to the direction of the bond $\langle ij \rangle$.
Similarly we can define the symmetrised triplet fields
\begin{eqnarray}
        \mb{\Psi}_{p_x} &=& \mb{\Psi}_x - \mb{\Psi}_{-x}, \nonumber \\
        \mb{\Psi}_{p_y} &=& \mb{\Psi}_y - \mb{\Psi}_{-y}. 
\end{eqnarray}
In the limit of fields varying slowly in space and time these
correspond to the Ginzburg-Landau order parameters for (extended)
$s$-wave, $d_{x^2-y^2}$ and $p$-wave pairing, respectively.  The $s$-
and $d$- wave pairing fields are even under inversion symmetry about
lattice site ${\bf r}_i$, while the triplet fields are odd.  In the
notation of Ref. \cite{annett}, these fields are the order parameters
for superconductivity in, respectively, the $A_{1g}$ and $B_{1g}$ and
$E_u$ representations of the tetragonal point group $D_{4h}$.

\section{The Saddle Point Solutions}

The saddle points of the effective action generate
Hartree-Fock-Gor'kov mean field theory, where the Bose fields become
static and spatially uniform order parameters:
\begin{eqnarray}
\psi_{\alpha} &=& -\frac{1}{\beta N}\int^{\beta}_0\sum_{i}
                    \Big< B_{\alpha}({\bf r}_i,\tau) \Big>_f,
                    \qquad \alpha = s,d\\
\mb{\Psi}_{\alpha} &=& -\frac{1}{\beta N}\int^{\beta}_0\sum_{i}
                    \Big< {\bf B}_{\alpha}({\bf r}_i,\tau)\Big>_f,
                    \qquad \alpha = x,y
\end{eqnarray}
where $N$ is the number of sites and $\langle\ldots\rangle_f$ denotes
self-consistent averaging with respect to the fermionic part of the
action, ${\cal S}_f$.
\\ \\
The transition temperature for a given order parameter, in the absence
of any of the others, is given by the solution of
\begin{equation}
\label{tcrit}
1=\frac{V}{2}\sum_{\epsilon}\frac{N^{\alpha}(\epsilon)}{\epsilon-\mu}
                  \tanh\left(\frac{\epsilon-\mu}{2T^{\alpha}_c}\right),
\end{equation}
in which the weighted density of states, $N^{\alpha}(\epsilon)$, is
\begin{equation}
N^{\alpha}=\frac{1}{N}\sum_{{\bf k}}
                \zeta^{\alpha}({\bf k})\zeta^{\alpha}({\bf k})
                \delta(\epsilon - \epsilon_{{\bf k}}).
\end{equation}
The form factors reflect the point group symmetries of the order
parameters:
\begin{equation}
      \zeta^{\alpha}({\bf k})=
        \left\{\begin{array}{cl}
        \cos(k_x)+\cos(k_y) & \alpha = s,\\
        \cos(k_x)-\cos(k_y) & \alpha = d,\\
        \sin(k_x)           & \alpha = {\bf p}_x,\\
        \sin(k_y)           & \alpha = {\bf p}_y.
        \end{array}\right.
\end{equation}
The solutions of Eq. \ref{tcrit} are shown in Fig. 1 for three values
of V.  The $s$-wave solution dominates for small and large fillings
with $d$-wave dominant near the van Hove peak at the centre of the
band.  The $p$-wave solutions are sub-dominant
everywhere. It is expected that interaction with the large $d$-wave
order parameter will suppress the $p$-wave T$_c$ even further.  It is
clear from this that $p$-wave pairing is irrelevant
in the bulk superconductor, and 
henceforth we ignore it and concentrate on $s$- and $d$-wave pairs
only.

\section{Beyond the Saddle Point: $s$- and $d$-wave Mixing}

Starting with the effective action, ${\cal S}_b$, we Fourier transform
in space and (imaginary) time and integrate out the fermions to give a
purely bosonic action,
\begin{eqnarray}
{\cal S}_b &=& 
    -\frac{V}{2}\sum_{q,\alpha}|\psi_{\alpha}(q)|^2 + {\rm Tr}\ln({\bf 1} - 
{\bf VG}_0),
\end{eqnarray}
where $q\equiv$({\bf q},$i\omega$), and the trace is over both
space-time and spinor indices.  The fermions of the original theory
live on in the form of the Nambu Green's function matrix,
\begin{eqnarray}
{\bf G}_0(k,k') \equiv 
   \left(\begin{array}{cc}
        G_0(k) & 0\\
        0 & -G_0^*(k)
   \end{array}\right)\delta_{kk'},
\end{eqnarray}
in which $G_0(k)=(i\omega_n-\epsilon_{\bf k}+\mu)^{-1}$ is the
Green's function for non-interacting fermions.  The interaction of
fermions and bosons occurs through the potential matrix,
\begin{eqnarray}
{\bf V}(k,k') \equiv 
    \frac{V}{\sqrt{2\beta N}}
        \left(
        \begin{array}{cc}
            0 & \psi_{\alpha}(k-k')\\
            \psi^*_{\alpha}(-k+k') & 0
        \end{array}
        \right)\zeta^{\alpha}_{{\bf k},{\bf k}'},
\end{eqnarray}
where the Einstein summation convention has been used for repeated
Greek indices and,
\begin{eqnarray}
\zeta^{\alpha}_{{\bf k},{\bf k}'} \equiv
        \frac{1}{2}\Big(\cos(k_x)+\cos(k'_x)\Big)
    \pm \frac{1}{2}\Big(\cos(k_y)+\cos(k'_y)\Big),\quad \alpha = s,d.
\end{eqnarray}
When $\psi_{\alpha}$ is constant in real space (i.e. ${\bf k}={\bf
k}'$), this takes a particularly simple form,
\begin{eqnarray}
\zeta^{\alpha}({\bf k}) \equiv \zeta^{\alpha}_{{\bf k},{\bf k}} 
                        = \cos(k_x)\pm\cos(k_y),\quad \alpha=s,d
\end{eqnarray}
as seen in the weighted densities of states.
\\ \\
Near T$_c$, where the $\psi_{\alpha}$ are small, we expand the
logarithm as a power series up to fourth order;
\begin{eqnarray}
    {\rm Tr}\ln({\bf 1} - {\bf VG}_0) &=&
    -\sum_{m=1}^4 \frac{1}{m}{\rm Tr}\left({\bf VG}_0\right)^m
    +{\cal O}\left(\psi^5\right).
 \end{eqnarray}
The odd terms in the series are exactly zero but the even terms
survive.  The quadratic contribution is
\begin{eqnarray}
{\rm Tr}\left({\bf VG}_0\right)^2
 = V^2\sum_{q}\psi^*_{\alpha}(q)\psi_{\beta}(q)\chi^{\alpha\beta}(q),
\end{eqnarray}
where the susceptibility is:
\begin{eqnarray}
\chi^{\alpha\beta}(q)
 &=& -\frac{1}{\beta N}\sum_{k}G_0(k)G_0^*(k+q)
    \zeta^{\alpha}_{{\bf k},{\bf k}+{\bf q}}
    \zeta^{\beta}_{{\bf k}+{\bf q},{\bf k}},\\
 &=& \frac{1}{N}\sum_{{\bf k}}
   \left\{
   \frac{f(\epsilon_{{\bf k}+{\bf q}}-\mu)
        +f(\epsilon_{{\bf k}}-\mu)-1}
        {i \omega_{\nu}+\epsilon_{{\bf k}+{\bf q}}
                                      +\epsilon_{{\bf k}}-2\mu}
   \right\}
    \zeta^{\alpha}_{{\bf k},{\bf k}+{\bf q}}
    \zeta^{\beta}_{{\bf k}+{\bf q},{\bf k}}.\nonumber
\end{eqnarray}
The quartic term is
\begin{eqnarray}
{\rm Tr}\left({\bf VG}_0\right)^4
 &=& \frac{V^4}{2\beta N}\sum_{\{q\}}
   \psi^*_{\alpha}(q)\psi_{\beta}(q')
   \psi^*_{\gamma}(q'')\psi_{\delta}(q-q'+q'')
   \chi^{\alpha\beta\gamma\delta}\left(q,q',q''\right).
   \nonumber \\
 &\;& 
\end{eqnarray}
It is sufficient to evaluate the four body susceptibility at
$q=q'=q''=0$ and then to treat it as a constant, $\chi \equiv \chi(0,0,0)$:
\begin{eqnarray}
 \chi^{\alpha\beta\gamma\delta}
 &=& \frac{1}{\beta N}\sum_{k} |G_0(k)|^4
   \zeta^{\alpha}({\bf k})\zeta^{\beta}({\bf k})
   \zeta^{\gamma}({\bf k})\zeta^{\delta}({\bf k}),\\
 &=& -\frac{1}{4N}\sum_{{\bf k}}
   \frac{1}{\xi}\frac{d}{d\xi}
        \left(
        \frac{\tanh\left(\beta\xi/2\right)}{\xi}
        \right)
   \zeta^{\alpha}({\bf k})\zeta^{\beta}({\bf k})
   \zeta^{\gamma}({\bf k})\zeta^{\delta}({\bf k}),\nonumber
\end{eqnarray}
where $\xi\equiv\epsilon_{{\bf k}}-\mu$.
\\ \\
Thus, after a trivial rescaling of $\psi$ by $V^{\frac{1}{2}}$, the
bosonic action to fourth order reads:
\begin{eqnarray}
{\cal S}_b &\approx& 
    -\frac{1}{2}\sum_{q,\alpha\beta}
      \left(
         \delta^{\alpha\beta}+V\chi^{\alpha\beta}(q)
      \right)
      \psi^*_{\alpha}(q)\psi_{\beta}(q)\\
           &\quad&\quad
     -\frac{V^2}{8\beta N}\chi^{\alpha\beta\gamma\delta}\sum_{\{q\}}
      \psi^*_{\alpha}(q)\psi_{\beta}(q')
      \psi^*_{\gamma}(q'')\psi_{\delta}(q-q'+q'').\nonumber
\end{eqnarray}
In order to derive a Landau-Ginzburg-Wilson functional from ${\cal
S}_b$, we expand to lowest order in small $i\omega_{\nu}$ and $|{\bf
q}|$:
\begin{eqnarray}
\frac{1}{2}
      \left(
         \delta^{\alpha\beta}+V\chi^{\alpha\beta}(q)
      \right)
&\simeq&
a^{\alpha\beta}
-id^{\alpha\beta}\omega_{\nu}
+\sum_{\mu=x,y}\frac{q^2_{\mu}}{2m^{\mu}_{\alpha\beta}},
\end{eqnarray}
where:
\begin{eqnarray}
a^{\alpha\beta} &=&
\frac{1}{2}
      \left(
         \delta^{\alpha\beta}-\frac{V}{2N}\sum_{{\bf k}}
        \left(
         \frac{\tanh\left(\beta\xi/2\right)}{\xi}
        \right)
         \zeta^{\alpha}({\bf k})\zeta^{\beta}({\bf k})
      \right),\\
d^{\alpha\beta} &=&
-\frac{V}{8N}\sum_{{\bf k}}
        \left(
         \frac{\tanh\left(\beta\xi/2\right)}{\xi^2}
        \right)
         \zeta^{\alpha}({\bf k})\zeta^{\beta}({\bf k}),\\
\label{mass}
\frac{1}{2m^{\mu}_{\alpha\beta}} &=&
-\frac{V}{16N}\sum_{{\bf k}}
            \Bigg\{
                \left(\frac{\partial^2\epsilon_{\bf k}}
                       {\partial k_{\mu}^2}\right)
                    \frac{d}{d\xi}  \left(
                       \frac{\tanh\left(\beta\xi/2\right)}{\xi}
                                    \right)
                    \zeta^{\alpha}({\bf k})\zeta^{\beta}({\bf k})\\
&\;&\qquad           +    \left(\frac{\partial\epsilon_{\bf k}}
                                     {\partial k_{\mu}}\right)^2
                    \left(
                        \frac{d^2}{d\xi^2}+\frac{1}{\xi}\frac{d}{d\xi} 
                    \right)
                    \left(
                        \frac{\tanh\left(\beta\xi/2\right)}{\xi}
                    \right)\zeta^{\alpha}({\bf k})\zeta^{\beta}({\bf k})
                    \nonumber\\
&\;&\qquad           +2 \left(\frac{\partial\epsilon_{\bf k}}
                                    {\partial k_{\mu}}\right)
                    \frac{d}{d\xi}  \left(
                       \frac{\tanh\left(\beta\xi/2\right)}{\xi}
                                    \right)
                       \left.\frac{\partial
                           (\zeta^{\alpha}_{{\bf k},{\bf k}+{\bf q}}
                            \zeta^{\beta}_{{\bf k}+{\bf q},{\bf k}})}
                                    {\partial q_{\mu}}\right|_{q=0}
                    \nonumber\\
&\;&\qquad           +2\left(
                       \frac{\tanh\left(\beta\xi/2\right)}{\xi}
                                    \right)
                       \left.\frac{\partial^2
                       (\zeta^{\alpha}_{{\bf k},{\bf k}+{\bf q}}
                        \zeta^{\beta}_{{\bf k}+{\bf q},{\bf k}})}
                        {\partial q_{\mu}^2}\right|_{q=0}
              \Bigg\}
\nonumber
\end{eqnarray}
where $\xi\equiv\epsilon_{{\bf k}}-\mu$.  The last two terms in
Eq. \ref{mass} are due to the ${\bf q}$ dependence of
$\zeta^{\alpha}$.
\\ \\
The resulting LGW functional is shown below expressed in real space:
\begin{equation}
{\cal S}_{LGW} 
= -\int^{\beta}_0 d\tau \sum_{\bf r}
\left[
    {\cal K} + {\cal V}
\right],
\end{equation}
where ${\cal K}$ is the `kinetic' part and ${\cal V}$ the `potential'
part of the action.  They are given by,
\begin{eqnarray}
%\end{eqnarray}
\label{eqkinetic}
{\cal K} 
&=&
d^{ss}\psi_s^* \partial_{\tau} \psi_s + \frac{|\nabla\psi_s|^2}{2m_{ss}}
+ d^{dd}\psi_d^* \partial_{\tau} \psi_d + \frac{|\nabla\psi_d|^2}{2m_{dd}}\\
&\;&\qquad\qquad\qquad\qquad
   +\;\frac{1}{2m_{sd}}
        \left(
        \psi^*_s(\nabla^2_x-\nabla^2_y)\psi_d + C.C.
        \right),\nonumber \\
\label{eqpotential}
{\cal V} 
&=&
a^{ss}|\psi_s|^2
   +a^{dd}|\psi_d|^2
   +b_s|\psi_s|^4
   +b_d|\psi_d|^4 \\
&\;&\qquad\qquad\qquad\qquad
   +\;\kappa|\psi_s|^2|\psi_d|^2
   +\frac{\kappa}{4}\left(
        \psi_s\psi_s
        \psi^*_d\psi^*_d
        +C.C.\right),\nonumber
\end{eqnarray}
where,
\begin{equation}
b_{s}  =  \frac{V^2}{8}\chi^{ssss},\quad\quad
b_{d}  =  \frac{V^2}{8}\chi^{dddd},\quad\quad
\kappa =  \frac{V^2}{2}\chi^{ssdd}.\\
\end{equation}
In the static and spatially uniform limit, ${\cal V}$ may be
interpreted as the Landau free energy.  We will examine the phases
arising from this before considering further the effects of
fluctuations. The free energy given in Eqs. \ref{eqkinetic}
\ref{eqpotential}  is of the form found earlier by Joynt\cite{joynt}
and by Feder and Kallin\cite{kallin}.

\section{Landau Theory}

\begin{figure}
\centerline{\epsfig{file=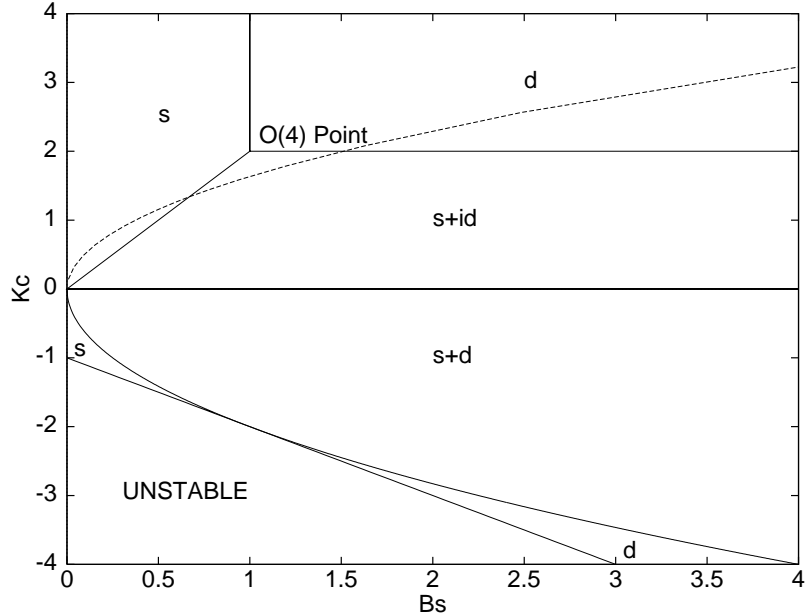,width=8cm,angle=-90}}
\caption{Landau stability plot for T$<$T$_c^s$,T$_c^d$.  In the vicinity of the bicritical point, the dashed line shows the transition from $d$-wave to $s$-wave via a mixed $s \pm id$ state as $\mu$ increases (right to left).}
\label{figure2}
\end{figure}

The phase diagram generated by ${\cal V}$ is found by simultaneously
solving,
\begin{equation}
\frac{\partial {\cal V}}{\partial \psi_s}
= \frac{\partial {\cal V}}{\partial \psi_d}
= 0.
\end{equation}
When $\kappa=0$, i.e. there is no coupling between the $s$- and
$d$-wave order parameters, solving $a^{ss}=0$ and $a^{dd}=0$ as 
functions of $\mu$ gives the critical temperature for each.  This is
equivalent to solving the linearised gap equation, Eq. \ref{tcrit}.
\\ \\
For non-zero $\kappa$ it is possible to have mixed phases.  To
determine the stable phase, two parameters are needed.  These are:
\begin{eqnarray}
B_s &=& \frac{b_s}{b_d}\left|\frac{a^{dd}}{a^{ss}}\right|^2,\\
K_c &=& \frac{\kappa(1+\frac{1}{2}\cos(2\Delta\theta)}{b_d}
        \left|\frac{a^{dd}}{a^{ss}}\right|,
\end{eqnarray}
where $\Delta\theta$ is the phase difference between $\psi_s$ and $\psi_d$.
Fig. \ref{figure2} shows the stable superconducting phase 
as a function of these two dimensionless parameters.   
Note that in Fig. \ref{figure2} the definition of $K_c$ is different
for $\kappa>0$, where $(\Delta\theta = \pi/2)$, and $\kappa<0$, where
$(\Delta\theta= 0)$.
\\ \\
As Fig. \ref{figure2} shows, depending on the parameter values $s$,
$d$, $s \pm d$ or $s \pm id$ phases are possible. The transition from
the pure $s$ to the pure $d$ phase can occur in a number of ways.  For
$K_c>2$, a single first-order transition separates the two phases,
analogous to a `spin flop' transition.  In this case, the $O(4)$ point
(corresponding to $B_s=1,K_c=2$) represents the bicritical point where
the line of first-order transitions $s\rightarrow d$ meets the two
normal $\rightarrow$ superconducting second-order lines.  When $K_c<2$
the transition occurs via two second-order phase transitions with an
intermediate mixed symmetry state, either $s \pm d$ or $s \pm id$.
The $O(4)$ point here represents the meeting of the four second-order
lines at a tetracritical point.
\\ \\
Using parameters derived from the nearest neighbour attractive Hubbard
model, the dashed line in Fig. \ref{figure2} shows the evolution of
($B_s$,$K_c$) for $V=2$ as $\mu$ changes in the region near the
crossing of T$_c^s$ and T$_c^d$.  As $\mu$ increases we move from the
far right hand side of the figure, where T$=$T$_c^s<$T$_c^d$ and
$d$-wave is dominant, to the far left, where T$=$T$_c^d<$T$_c^s$ and
$s$-wave is dominant.  In these extremes, the dominant $d$-wave
($s$-wave) order parameter suppresses the sub-dominant $s$-wave
($d$-wave) one even though T$<$T$_c^s$(T$_c^d$). Eventually, however,
a mixed $s \pm id$ phase arises.  This behaviour is seen for all $V$. In
particular the mixed $s \pm d$ phase is never realised for any values of
$B_s$ and $K_c$ derived from the nearest neighbour attractive Hubbard
model unless an orthorhombic distortion is introduced which is beyond
the scope of this paper.

\section{The Large $V$ Limit}

In the large $V$ limit it is necessary to retain the
thermal and quantum fluctuations in the Bose fields derived in section 4.
The superconducting phase transition then becomes the
(Kosterlitz-Thouless in 2-dimensions) Bose condensation
of these quantum fields. A detailed analysis is beyond the scope
of the present paper, but it is possible to draw some
qualitative conclusions based on the effective action derived in 
sections 3 and  5.  
\\ \\
Fig. \ref{figure1} shows the evolution of the T$_c$ for
$s$, $d$ and $p$-wave pairing as a function of
$\mu$ for various values of $V/t$.  The small $V$ limit,
represented by Fig. 1(a), corresponds to the BCS weak coupling limit
in which the pairing is a small perturbation on the filled Fermi sea.
On the other hand as $V$ is increased (Fig 1(b)-(c)) the 
behaviour changes markedly.  One can see that the 
$s$-wave T$_c$ becomes significant even at $\mu=0$, and that the
$s$ and $d$-wave curves become increasingly similar as
$V$ is increased.  This has a simple physical interpretation:
in the large $V$ limit the pairs become nearly localised on bonds,
and the most favourable  state is a singlet pair (stable
compared to triplet by an energy of order $t^2/V$). 
As $V$ increases the effective hopping for pairs decreases
(also as $t^2/V$) and so $s$- and $d$-wave pairs become
nearly degenerate.  This degeneracy implies that the Bose
condensation of pairs in the large $V$ nearest-neighbour Hubbard model
is qualitatively different from that in the on-site attractive
Hubbard model.
\\ \\
Figures 1(b) and 1(c) also shows that for large $V$ superconductivity
occurs even when the chemical potential is below the bottom of the
electronic band at $-4t$.  For these values of $\mu$ the fermions can
only exist as bound pairs, and not as free fermions.  This is
analogous to the large $V$ on-site s-wave case discussed by Randeria,
Duan and Shieh \cite{randeria}, in which the criterion for Bose
condensation to occur in the low density limit was shown to be the
same as the criterion for formation of a two particle bound
state. Figure 1(c) implies that in the low density, $\mu<-4t$, limit
two-particle bound states are formed. An extended $s$-wave
superconducting state then occurs when these preformed pairs Bose
condense at temperatures below the mean-field T$_c$ shown in
Fig. 1(c).

\section{Conclusions}

We have derived an appropriate Landau-Ginzburg-Wilson effective action
to describe the nearest neighbour attractive Hubbard model.  This
allows us to discuss both the small $V$ BCS and the large $V$
Bose-Einstein condensation limits on the same footing, as was done for
the on-site attractive Hubbard model some years ago\cite{demelo}.  The
nearest neighbour Hubbard model is especially interesting because the
phase diagram includes regions of both (extended) $s$-wave
superconductivity and $d$-wave pairing.  We find that $p$-wave pairing
is never stable.  We have studied closely the region of the phase
diagram near the cross-over from $s$- to $d$-wave pairing, and find
that the two phases are always separated by two second-order phase
transitions with an intermediate phase of $s \pm id$ superconductivity.
\\ \\
In future it will be interesting to examine more carefully the large
$V$ Bose-Einstein condensation limit of this model, since the extended
$s$-wave and $d$-wave pairing states become nearly degenerate.  It
will also be interesting to see how closely features of this model,
such as the existence of $s$, $d$ and $s \pm id$ phases, correspond to the
actual experimental features of the cuprates.
\\ \\
This work was supported by EPSRC Grant No. GR/L22454,
and by the Office of Naval Research Grant No. N00014-95-1-0398.

\end{document}